\documentclass[letterpaper, aps, showpacs,twocolumn]{revtex4}
\usepackage{amsmath}
\usepackage{amssymb}
\usepackage{latexsym}
\usepackage[dvips]{graphics}
\usepackage{hyperref}

\begin{document}

\title{Anomalous quantum reflection as a quasi-dynamical damping effect}
\author{Alexander Jurisch and Jan-Michael Rost}
\affiliation{Max-Planck-Institut f\"ur Physik komplexer Systeme\\
N\"othnitzerstr. \!\!38, 01187 Dresden, Germany}
\begin{abstract}
We develop a quasi-analytical theory for the quantum reflection amplitude of Bose-Einstein condensates. We derive and calculate the decay-width of a Bose-Einstein condensate. A general relation between the time-dependent decay-law of the system and its quantum reflection amplitude allows us to explain the quantum reflection anomaly of Bose-Einstein condensates present in BEC-surface systems as a direct consequence of the repulsive particle interaction.
\end{abstract}
\maketitle
\section{Introduction}
In recent experiments the quantum reflection of a sodium BEC from a silicon surface has been investigated, \cite{Pas1,Pas2}. Surprisingly the quantum reflection probability fell off to zero at threshold and did not approach unity as expected from corresponding single atom beam theory e.g. \cite{CotFriTro, FriJacMei} and experiments e.g., \cite{Shi, DruDeK, ObeKouShiFujShi, ObeTasShiShi, KouObe}.

Theoretically so far, two numerical studies were able to qualitatively produce the quantum reflection anomaly. In \cite{Pas1, Pas2} a model was used to motivate the quantum reflection anomaly by introducing an empirical potential which modifies the atom-surface interaction. This lead qualitatively to the observed effect. In the second numerical study \cite{ScoMarFroShe} an ab initio numerical simulation of the experimental situation reported in \cite{Pas1} in three dimensions was used to explain the anomalous behaviour of the quantum reflection probability with the formation of scattering halos and the building of vortices that destroy the coherence of the condensate while being reflected. The anomalous behaviour of the reflection probability emerged when taking into account only regions of the reflected BEC with more than 25\% of its initial density.

In the following we present a quasi-analytical approach to describe the anomalous behaviour of the quantum reflection probability $|R|^{2}$. Due to its simplicity the origin of the anomaly is elucidated. Furthermore, we identify scattering along the axis of normal incidence on the surface as being mainly responsible for the anomalous behaviour of the quantum reflection probability. On base of this, we may use the framework of the well-understood spherical quantum reflection trap model \cite{JurFri1, JurRos}. A spherical model has the advantage that there is only the normal direction for s-waves. This property strongly reduces the technical complexity of the general three-dimensional system and allows us to focus on the physics behind the quantum reflection of a BEC. Furthermore, our model relies solely on the presence of particle-interaction and mean-field dynamics.

\section{Quantum Reflection}
For monochromatic atomic beams or linear wave-packets the universal law of quantum reflection is obtained by the fact that atom-surface potentials behave as step-potentials for incident momenta close to threshold, see \cite{CotFriTro, FriJacMei, JurFri1, MadFri}. The universal threshold-law of quantum reflection is given by
\begin{equation}
\lim_{k\,\rightarrow\,0}\,\left|R(k)\right|\,=\,1\,-\,2\,\,b\,k\quad,
\label{universalthresholdlaw}\end{equation}
where the parameter $b$ is called the threshold-length and $k$ is the incident momentum. In case of an atom-surface potential it has been shown \cite{FriJacMei, JurFri2, DruDeK} that close to threshold the retarded Casimir-Polder tail of the potential dominates the process of quantum reflection. The retarded Casimir-Polder tail is given by
\begin{equation}
U(z)\,=\,-\,\frac{\hbar^{2}}{2\,m}\frac{\beta_{4}^{2}}{\left|z\,-\,L\right|^{4}}\quad,
\label{CasimirPolder}\end{equation}
where $L$ denotes the location of the surface. For low incident momenta such potentials behave like
\begin{equation}
U(z)\,=\,-\,\frac{\hbar^{2}}{2\,m}\,b^{-2}\,\theta(z\,-\,L),\quad b\,=\,\beta_{4}\quad.
\label{steppotential}\end{equation}

For the quantum reflection of linear wave-packets it has been shown in \cite{JurFri1} that the quantum reflection amplitude $|R(k)|$ can be extended to the whole $k$-space by
\begin{equation}
\left|R(k)\right|\,=\,1\,-\,2\,\,b\,k\,\approx\,\exp[-2\,b\,k]\quad,
\label{extendedthresholdlaw}\end{equation}
with reasonable accuracy. It further has been shown that the decay-law for a linear wave-packet is linked to Eq.(\ref{extendedthresholdlaw}) by
\begin{equation}
P(k, t)\,=\,|R(k)|^{\frac{2\,k\,t}{2\,m\,L}}\,=\,\exp\left[-4\,\frac{b\,k^{2}\,t}{2\,m\,L}\right]\quad.
\label{lineardecaylaw}\end{equation}

\section{Derivation of the condensate's damping function}
To study the influence of particle-interaction on the threshold-behaviour of quantum reflection, we turn to the radial system that we have examined in detail from a dynamical point of view, \cite{JurRos}.
The radial Gross-Pitaevskii equation that describes the time evolution of an initial state $\Psi(x,\tau\,=\,0)$ in the atom-surface system in scaled form is given by
\begin{eqnarray}
i\,\frac{\partial}{\partial\,\tau}\,\psi(x,\tau)\,&=&\,-\frac{\partial^{2}}{\partial x^{2}}\psi(x,t)\,-\,\sigma^{2}\theta\left(x-1\right)\,\psi(x,\tau)\,\nonumber\\
&+&\,\gamma\frac{\left|\psi(x,\tau)\right|^2}{x^{2}}\,\psi(x,\tau)\quad.
\label{GP}\end{eqnarray}
Scaling is carried out by the intrinsic length of the system $L$,
\begin{eqnarray}
x\,=\,\frac{r}{L},\quad\kappa\,=\,k\,L,\quad\sigma\,=\,\frac{L}{\beta_{4}}\nonumber\\
\gamma\,=\,\frac{2\,a_{{\rm{int}}}}{L}\,N,\quad\tau\,=\,\frac{t\,\hbar}{2\,m\,L^{2}}\quad.
\end{eqnarray}
In \cite{JurRos} we have shown that a repulsive self-interaction provides an additional contribution to the total energy of the system that is responsible for a faster decay of the particle-density than it is found in the linear case. Dynamically, this is explained by a transformation of the self-interaction energy into kinetic energy as the initial state evoles in time. In $k$-space, we interpret the presence of the self-interaction as an additional barrier. The $k$-modes close to threshold lie at the base of this barrier and are thus damped away by driving them over the edge of the potential. The effect of the self-interaction is strongest at the beginning of the time-evolution, because its magnitude is directly related to the incident particle-density. For larger times, a considerable fraction of particle-density has already decayed, such that the influence of the self-interaction becomes more and more negligible.

For small times, we assume that the decay of the system can be described by a product-ansatz
\begin{eqnarray}
P(\kappa,\tau;\sigma,\gamma)\,&=&\,P_{\sigma}(\kappa,\tau)\,P_{\gamma}(\kappa,\tau)\,=\,\left|R_{\sigma}(\kappa)\,R_{\gamma}(\kappa)\right|^{2\,\kappa\,\tau}\nonumber\\
&=&\,\exp\left[-4\,\frac{\kappa^{2}}{\sigma}\,\tau\right]\left|R_{\gamma}(\kappa)\right|^{2\,\kappa\,\tau}\quad.
\label{productansatz}\end{eqnarray}
In Eq. (\ref{productansatz}) the additional damping induced by the particle-interaction is described by $P_{\gamma}(\kappa, \tau)$. The product-ansatz of Eq. (\ref{productansatz}) reflects our assumption that for early times the effect of the step-potential and the particle-interaction can be treated independently. By this assumption, we derive the additional damping induced by the particle-interaction without taking into account the presence of the atom-surface potential in Eq. (\ref{GP}).

The wave-function $\psi$ that solves Eq. (\ref{GP}) without potential can be decomposed into its momentum components by setting
\begin{eqnarray}
&&\psi(x,\tau)=\int d\kappa A(\kappa)\varphi_{\kappa}(x)\exp[-i\kappa^{2}\tau]\phi_{\gamma}(\kappa,\tau),\nonumber\\
&&\phi_{\gamma}(\kappa,\tau)\,=\,\exp\left[-i\,\int_{0}^{\tau}\,d\tau'\,E_{\gamma}(\kappa, \tau')\right]\quad.
\label{fouriertransform1}\end{eqnarray}
All effects of particle-interaction in Eq. (\ref{fouriertransform1}) are described by $\phi$.
The Fourier-decomposition in Eq. (\ref{fouriertransform1}) is defined by
\begin{equation}
A(\kappa)\,=\,\int_{0}^{1}\,dx\,\Psi(x,\tau\,=\,0)\,\varphi_{\kappa}(x)\quad,
\label{fouriertransform2}\end{equation}
and the basis of the system is
\begin{equation}
\varphi_{\kappa}(x)=\sqrt{\frac{2}{\pi}}\,\sin[\kappa x]\quad.
\label{basemodes1}\end{equation}

The decay can be calculated from the time-dependent dispersion, that is obtained by inserting the Fourier-decomposition Eq. (\ref{fouriertransform1}) in Eq. (\ref{GP}), neglecting the atom-surface potential, multiplying by $\varphi$ from the left, integrating out $x$ and separating off the non-interacting parts. We thus obtain the time-dependent dispersion of the self-interacting system
\begin{eqnarray}
&&i\frac{\partial_{\tau}\,\phi_{\gamma}(\kappa,\tau)}{\phi_{\gamma}(\kappa,\tau)}\,=\,E_{\gamma}(\kappa,\tau)\,=\,\gamma\,\int_{0}^{\infty}\,d\kappa_{1}\,d\kappa_{2}\nonumber\\
&&\times\,V(\kappa, \kappa_{1},\kappa_{2})\,
\phi^{*}_{\gamma}(\kappa_{2},\tau)\,\phi_{\gamma}(\kappa_{1},\tau)\nonumber\\
&&\times\,\exp\left[-i\,(\kappa_{1}^{2}-\kappa_{2}^{2})\,\tau\right]\quad.
\label{timedependentdispersion}\end{eqnarray}
The function $V(\kappa, \kappa_{1},\kappa_{2})$ in Eq. (\ref{timedependentdispersion}) is a vertex-function, that describes the interaction of the modes of the system in momentum space. The main contribution to the vertex stems from the threshold region. We find 
\begin{eqnarray}
V(\kappa, \kappa_{1},\kappa_{2})\,=\,3\,\kappa^{-2}\,A(\kappa_{1})\,A(\kappa_{2})\,\frac{2}{\pi}\times\nonumber\\
\int_{0}^{1}\,dx\,\sin^{2}[\kappa\,x]\,\frac{\sin[\kappa_{1}\,x]\,\sin[\kappa_{2}\,x]}{x^{2}}\quad.
\label{vertexfunction}\end{eqnarray}
The decay of the system is described by the imaginary part of the dispersion, that can be identified by the line-width $\Gamma_{\gamma}$. To calculate $\Gamma_{\gamma}$, we expand $E_{\gamma}(\kappa,\tau),\,\phi_{\gamma}(\kappa,\tau)$ in frequency-space, obtaining
\begin{eqnarray}
E_{\gamma}(\kappa,\tau)\,&=&\,\int\,\frac{d\,\omega}{2\,\pi}\,E_{\gamma}(\kappa,\omega)\,\exp[-i\,\omega\,\,\tau]\,\quad,\nonumber\\
\phi_{\gamma}(\kappa,\tau)\,&=&\,\int\,\frac{d\,\omega}{2\,\pi}\,\phi_{\gamma}(\kappa,\omega)\,\exp[-i\,\omega\,\,\tau]\,\quad.
\label{frequencyspacedecomposition}\end{eqnarray}
By using Eq. (\ref{frequencyspacedecomposition}) we can perform a Laplace-transform on Eq. (\ref{timedependentdispersion}). The imaginary part we are interested in can be extracted by using the well-known relation $\frac{1}{\Omega-\omega+i\,\epsilon}\,=\,P\frac{1}{\Omega-\omega}\,-\,i\,\pi\,\delta(\Omega-\omega)$. We integrate out the $\delta$-function and find
\begin{eqnarray}
&-&\Im{E_{\gamma}(\kappa, \Omega)}\,=\,\Gamma_{\gamma}(\kappa, \Omega)\,\nonumber\\
&=&\,\gamma\,\int_{0}^{\infty}\,d\kappa_{1}\,d\kappa_{2}\,V(\kappa, \kappa_{1},\kappa_{2})\,\times\,\nonumber\\
&&\int\,\frac{d\omega_{2}}{2\pi}\phi^{*}_{\gamma}(\kappa_{2},\omega_{2})\phi_{\gamma}(\kappa_{1},\Omega-\kappa_{1}^{2}+\kappa_{2}^{2}+\omega_{2})\quad.
\label{linewidth1}\end{eqnarray}
So far, we have only neglected the atom-surface potential. Since we are interested in the damping that is induced by the initial state we now make a relaxational ansatz that treats $\Gamma_{\gamma}$ independent of $\omega$ by setting
\begin{equation}
\phi_{\gamma}(\kappa, \omega) \approx \frac{i}{\omega\,+\,i\,\Gamma_{\gamma}(\kappa)}\quad.
\label{ansatzlinewidth1}\end{equation}
We insert Eq. (\ref{ansatzlinewidth1}) into Eq. (\ref{linewidth1}), carry out the remaining integration by the residual theorem and arrive at
\begin{eqnarray}
&&\Gamma_{\gamma}(\kappa)\,=\,i\,\gamma\,\int_{0}^{\infty}\,d\kappa_{1}\,d\kappa_{2}\times\nonumber\\
&&\frac{V(\kappa,\kappa_{1},\kappa_{2})}{\kappa_{2}^{2}\,-\,\kappa_{1}^{2}\,+\,i\,\left(\Gamma_{\gamma}(\kappa_{1})\,+\,\Gamma_{\gamma}(\kappa_{2})\right)}\quad. 
\label{linewidthequation}\end{eqnarray}
Equation (\ref{linewidthequation}) is a self-consistent equation for the line-width of the self-interacting part of the system. From Eqs. (\ref{ansatzlinewidth1}, \ref{linewidthequation}) it follows, that the self-interacting part of the system initially decays like
\begin{equation}
\left|\phi_{\gamma}(\kappa, \tau)\right|^{2}\,\approx\,P_{\gamma}(\kappa, \tau)\,=\,\exp\left[-2\,\Gamma_{\gamma}(\kappa)\,\tau\right]\quad.
\label{approxdecay}\end{equation}
Together with Eq. (\ref{productansatz}) the quantum reflection amplitude readily follows as
\begin{eqnarray}
|R(\kappa;\sigma, \gamma)|\,&=&\,|R_{\sigma}(\kappa)\,R_{\gamma}(\kappa)|\nonumber\\
&=&\,\exp\left[-2\frac{\kappa}{\sigma}\right]\,\exp\left[-\frac{\Gamma_{\gamma}(\kappa)}{\kappa}\right]\quad. 
\label{reflectionamplitude}\end{eqnarray}
Following \cite{JurFri1} the quantum reflection amplitude Eq. (\ref{reflectionamplitude}) describes the decay of the initial state in a uniform and quasi-stationary way by
\begin{equation}
\left<\rho(\tau)\right>\,=\,\int_{0}^{\infty}\,d\kappa\,\left|A(\kappa)\right|^{2}\,|R(\kappa;\sigma, \gamma)|^{2\,\kappa\,\tau}\quad,
\label{quasistationarydecay}\end{equation}
where $\left<\rho(\tau)\right>$ in Eq. (\ref{quasistationarydecay}) stands for the mean-value of the time-dependent particle density $\rho(\tau)\,=\,\int_{0}^{1}\,dx\,\left|\psi(x, \tau)\right|^{2}$ that is obtained from numerics.

\section{Discussion}
Now we are going to test the fidelity of our theory by comparing it to numerical results. To make contact with our analysis in \cite{JurFri1, JurRos}, we use the initial state
\begin{equation}
\Psi(x,\tau\,=\,0)\,=\,\mathcal{N}\,x\,\exp\left[-\,a\,x\right]\theta\left[1\,-\,x\right]\quad,
\label{initialwavepacket}\end{equation}
where $\mathcal{N}$ is the normalization constant and $a$ is the diffuseness of the wave-packet. A diffuseness of $a\,=\,5$, gives an initial kinetic energy $E_{\rm{kin}} = 1.5\,\times\,10^{-15}$\,[a.u.] for sodium that corresponds to temperatures of approximatly  1\,nK, comparable to the experimental setup in \cite{Pas1, Pas2}. As in \cite{JurFri1} we chose a length-scale $L\,=\,4.47\,\times\,10^{5}$\,[a.u.], and together with a potential-strength $\beta_{4}\,=\,1.494\,\times\,10^{4}$\,[a.u.] and mass $m\,=\,4.22\,\times\,10^{5}$\,[a.u.] for sodium we obtain a scaled potential-strength $\sigma\,=\,30$. To illustrate the effect of particle-interaction, we chose $\gamma\,=\,0.5$\quad. 
\begin{figure}[t]\centering\vspace{-1.05cm}
\rotatebox{-90.0}{\scalebox{0.35}{\includegraphics{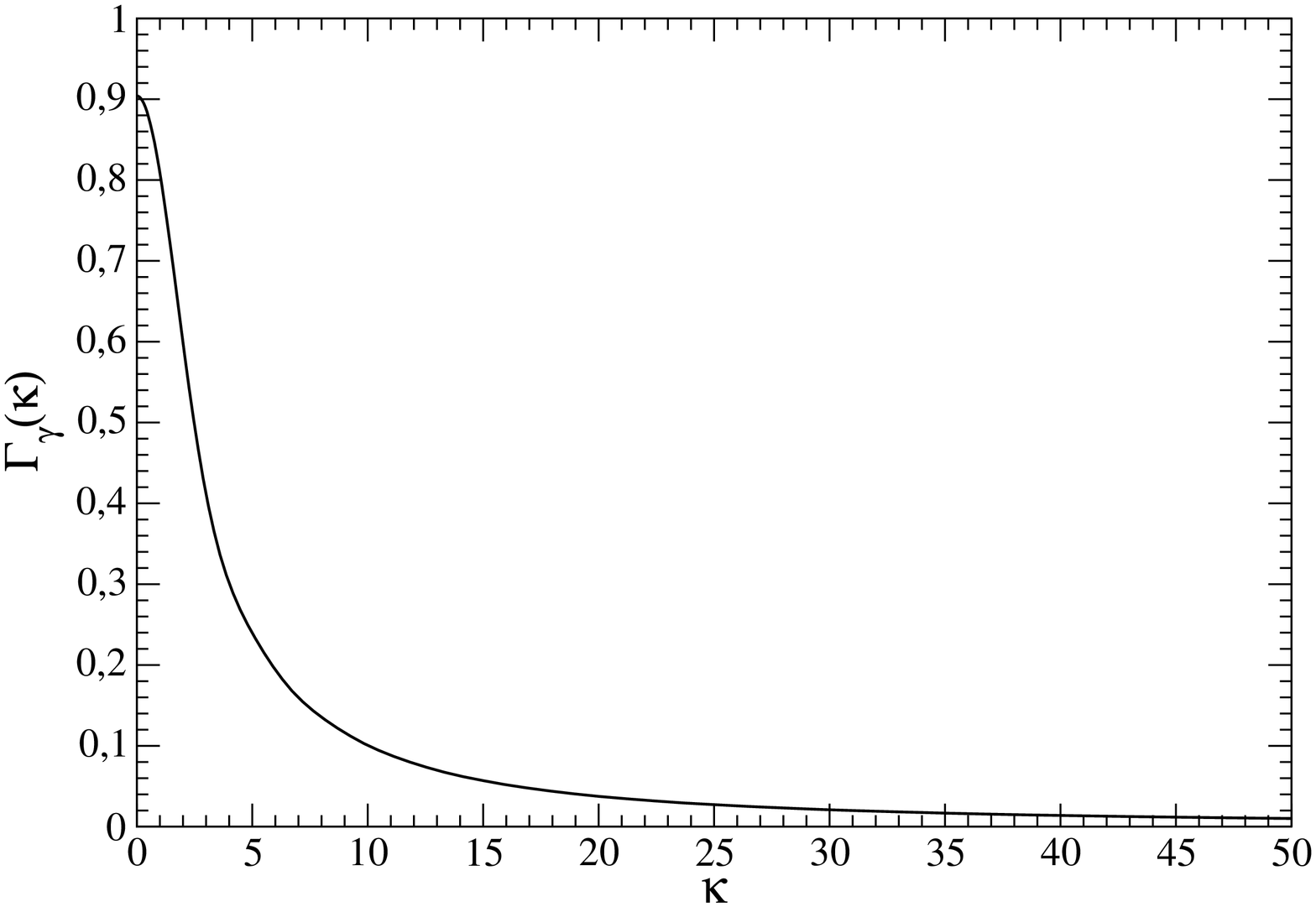}}}
\caption{\footnotesize{Damping-function $\Gamma_{\gamma}(\kappa)$ calculated by Eq. (\ref{linewidthequation}) for $\gamma\,=\,0.5$.}}
\label{fig1}\end{figure}

Figure (\ref{fig1}) shows the additional damping that is induced by the particle interaction. It is clearly visible that the low-lying $k$-modes are strongly damped, while higher $k$-modes experience only a moderate to negligible additional damping and thus show a regular decay according to universal quantum reflection Eqs. (\ref{extendedthresholdlaw}, \ref{lineardecaylaw}).

\begin{figure}[t]\centering\vspace{-1.05cm}
\rotatebox{-90.0}{\scalebox{0.35}{\includegraphics{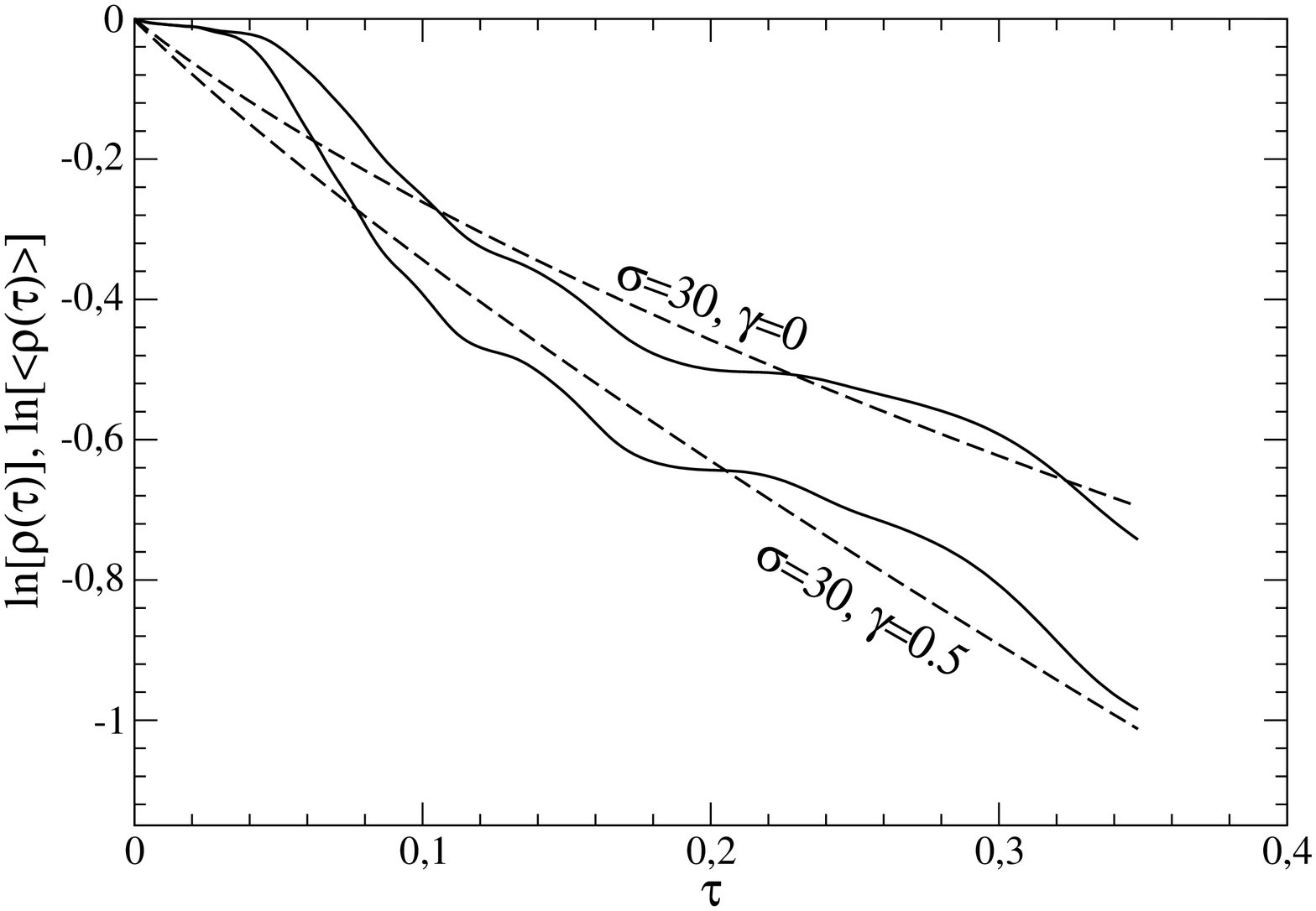}}}
\caption{\footnotesize{Comparison between numerically obtained decaying densities $\rho(\tau)$ (full lines) and approximated decaying densities $\left<\rho(\tau)\right>$ (dashed lines) by Eq. (\ref{approxdecay}) for ($\sigma\,=\,30,\,\gamma\,=\,0$) above and ($\sigma\,=\,30,\,\gamma\,=\,0.5$) below.}}
\label{fig2}\end{figure}

Figure (\ref{fig2}) shows a comparison between numerically calculated densities as function of the scaled time $\tau$ and their mean-value approximation according to Eq. (\ref{approxdecay}) for early times of the decay of the system. We have chosen a logarithmic ordinate for a better resolution. Figure (\ref{fig2}) shows clearly that after the einschwingvorgang is completed the mean-value approximation Eq. (\ref{approxdecay}) interpolates the numerical curves. The approximation Eq. (\ref{approxdecay}) describes the decay in a quasi-stationary and uniform way and thus does not reflect the oscillations of the numerically obtained data, that originate from the motion of the wave-packet on the spatial range of the step. For times $\tau\,\geq\,0.35$ our approach is not reliable anymore, because it takes only into account the damping according to the initial state, that naturally must overestimate the decay of the particle density for later times. A scaled time $\tau\,\sim\,0.35$ as a range of fidelity corresponds to times $t\,\sim\,0.144$ seconds.

\begin{figure}[t]\centering\vspace{-1.05cm}
\rotatebox{-90.0}{\scalebox{0.35}{\includegraphics{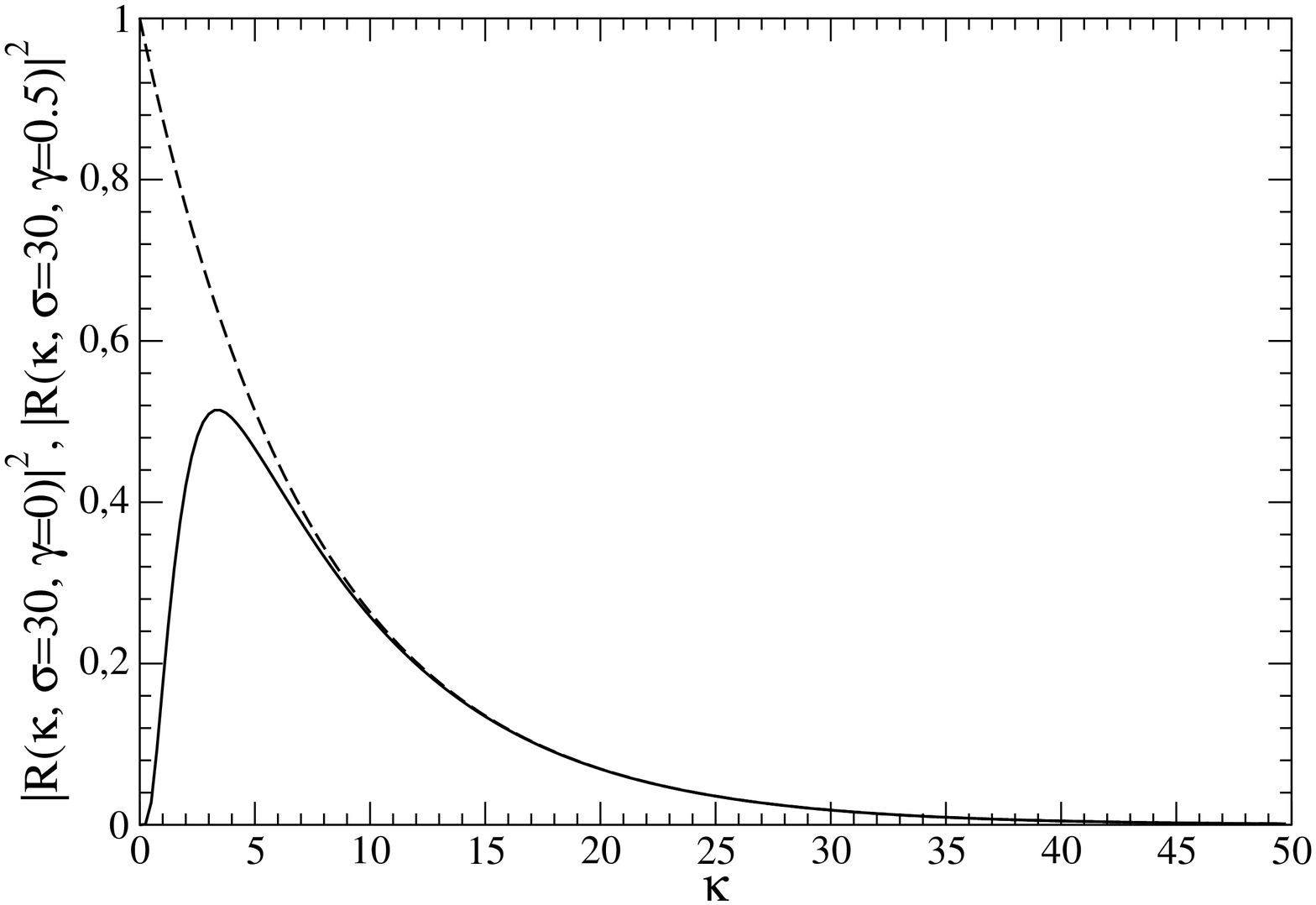}}}
\caption{\footnotesize{Quantum reflection probabilities according to Eq. (\ref{reflectionamplitude}) for ($\sigma\,=\,30,\,\gamma\,=\,0.5$) (full line) and ($\sigma\,=\,30,\,\gamma\,=\,0$) (dashed line).}}
\label{fig3}\end{figure}

Figure (\ref{fig3}) shows the quantum reflection probabilities calculated according to Eq. (\ref{reflectionamplitude}). Figure (\ref{fig3}) clearly demonstrates how the additional damping that is induced by the particle-interaction influences the quantum reflection probability. The depletion of the quantum reflection probability for low-lying $k$-modes corresponds to the results reported in \cite{Pas1, Pas2, ScoMarFroShe}. The curves shown in Fig. (\ref{fig3}) must be interpreted as a quasi-stationary mean-value of the quantum reflection probability that is to be expected for times $\tau\,\leq\,0.35$. For later times, as the influence of particle-interaction declines along with the decaying density, we expect a reduction of the anomalous behaviour that, for very large times, will experience a cross-over to the universal law of quantum reflection Eqs. (\ref{universalthresholdlaw}, \ref{lineardecaylaw}).

\section{Summary and Conclusion}
Together with our investigations on the dynamical properties of a BEC-surface system, \cite{JurRos}, the behaviour of Eq. (\ref{reflectionamplitude}) as shown in Fig. (\ref{fig3}) proves, that quantum reflection has only a limited effect on BECs with repulsive particle-interaction. We thus explain the quantum reflection anomaly by the fact, that especially the low energetic, collective components of the wave-packet moving normal to the surface react extremely sensitively on the presence of an additional positive energetic contribution as a repulsive particle-interaction provides. The crucial role of the repulsive particle-interaction and its transformation into kinetic energy was already mentioned in \cite{Pas1} and theoretically demonstrated in \cite{JurRos}. The influence of the repulsive interaction on higher energetic components is by far less dramatic, explaining the regular behaviour of Eq. (\ref{reflectionamplitude}) for higher momenta. The fact that we can explain the quantum reflection anomaly with the behaviour of the line-width as function of momentum may give hints for future experiments to measure the line-width of the condensate instead of the reflection-probability, which, as emphazised in \cite{Pas1, Pas2}, is quite sensitive to external influences.

The purpose of our present theory is neither to exactly recalculate the experimental data from \cite{Pas1, Pas2}, nor the numerical results reported in \cite{ScoMarFroShe}, but to elucidate the physical mechanism that is responsible for the anomalous behaviour of the quantum reflection probability on the basis of a simple but well-understood model. Our quasi-analytical theory focuses exclusively on the direction of normal incidence, which we have assumed and proved to contribute significantly to the quantum reflection anomaly. Our analysis demonstrates that the anomaly exists and that it may be found in experiments. Furthermore, the experiments reported in \cite{Pas1, Pas2} so far contain enough uncertainties to doubt about the relevance of a direct and quantitative comparison between theory and experiment in the present state of the art. However, our analysis may encourage further experimental work in this field.

In the framework of the simple spherical model, \cite{JurRos}, we have developed a quasi-analytical theory for the anomalous behaviour of the quantum reflection probability. Our theory explains the anomalous behaviour of the quantum reflection probability of BECs as a direct consequence of the repulsive particle-interaction. The key-quantity for understanding the quantum reflection anomaly of BECs was shown to be the line-width of the condensate as function of momentum.

\end{document}